\documentclass[twocolumn,10pt]{tsfp}
\usepackage{flushend}
\usepackage{graphicx}
\usepackage[authoryear,round]{natbib}
\usepackage{fancyhdr}
\usepackage[per-mode=symbol]{siunitx}
\usepackage{subcaption}
\graphicspath{{Figures/}}
 
\title{Computational and Experimental Study of Spanwise Synthetic Jet Flow Control}

\author{Howard H. Ho
    \affiliation{
	Mechanical and Industrial Engineering\\
	University of Toronto\\
	Ontario, M5S 3G8, Canada\\
        howard.ho@mail.utoronto.ca
    }	
}

\author{Adnan Machado
    \affiliation{
        Mechanical and Industrial Engineering\\
	University of Toronto\\
	Ontario, M5S 3G8, Canada\\
        adnan.machado@mail.utoronto.ca
    }
}

\author{Pierre E. Sullivan
    \affiliation{
        Mechanical and Industrial Engineering \\
	University of Toronto\\
	Ontario, M5S 3G8, Canada\\
	pierre.sullivan@utoronto.ca
    }
}

\begin{document}

\maketitle   %Print title matter
\thispagestyle{fancy}

% Set the font to 9pt.
\fontsize{9}{11}\selectfont

%%%%%%%%%%%%%%%%%%%%%%%%%%%%%%%%%%%%%%%%%%%%%%%%%%%%%%%%%%%%%%%%%%%%%%
\section*{ABSTRACT}
This study investigates the reattachment of flow on a stalled NACA 0025 airfoil with an array of circular synthetic jet actuators. Experimental flow visualizations are used to assess the spanwise control authority of the array and the three-dimensionality of the flow. Numerical simulations provide insights into the flow structures created by the actuation, and how they evolve with different parameters.
%%%%%%%%%%%%%%%%%%%%%%%%%%%%%%%%%%%%%%%%%%%%%%%%%%%%%%%%%%%%%%%%%%%%%%
\section*{INTRODUCTION}
Flow control of reattaching separated flows has many applications, such as preventing stall in low-speed, high-altitude aircraft for surveying, and improving the aerodynamics of wind turbine blades with short chord lengths. Synthetic jet actuators (SJAs) are zero-net-mass-flux devices that add momentum to the flow via periodic ingestion and expulsion cycles. In contrast to continuous jets, SJAs do not require plumbing and reservoirs which makes them ideal for active flow control. The process of flow reattachment with low SJA actuation frequencies is primarily attributed to the presence of spanwise vortical structures that convect over the suction side of the airfoil. These structures entrain high-momentum fluid from the freestream into the separated shear layer \citep{Salunkhe2016, Rice2018, Xu2023}. This transported momentum energizes the shear layer, enabling the flow to overcome the adverse pressure gradient, leading to flow reattachment. High-frequency control was found to result in steadier aerodynamic effects \citep{Amitay2002,Xu2023,Machado2023}, and thus the flow structures associated with this control strategy are of interest. The majority of experimental research focuses on the effects of flow control at the symmetry plane of the wing. However, these techniques seldom assess the spanwise control authority, leaving a gap in the understanding of the three-dimensional controlled flow. We present a comprehensive study utilizing smoke visualization to understand the three-dimensionality of the flow, as well as a preliminary analysis of the effect of varying the blowing ratio, $C_B$. These results are augmented by computational fluid dynamics (CFD) which allows for detailed analysis of the underlying flow structures responsible for flow reattachment.

\section*{EXPERIMENTAL METHOD}
Experiments were conducted in the low-speed recirculating wind tunnel at the University of Toronto. The test section has dimensions of 5~\unit{\metre}~$\times$~0.91~\unit{\metre}~$\times$~1.22~\unit{\metre} and features acrylic windows on the top and side walls for observation and measurement. The wind tunnel can produce speeds between 3--18~\unit[per-mode = symbol]{\metre\per\second} with a turbulence intensity of less than 1\%. For the experiments conducted, the wind tunnel was operated at a freestream velocity of $U_\infty=5.1$~\unit[per-mode = symbol]{\metre\per\second}, resulting in a chord-based Reynolds number of $\mathrm{Re}_c=10^5$.

A NACA 0025 airfoil was placed in the wind tunnel with the leading edge approximately 40~\unit{\centi\metre} from the test section inlet. The aluminum wing has an aspect ratio of approximately 3, with a span of $b=885$~\unit{\milli\metre}, and a chord length of $c=300$~\unit{\milli\metre}. The wing comprises 3 parts, with a hollow center third to house the sensors and actuators. In the center, there is a \SI{317}{\milli\metre}~$\times$~\SI{58}{\milli\metre} rectangular cutout where the microblower array is installed, with a flush 0.8 mm hole for the nozzle of each SJA. The angle of attack was set to $\alpha=\SI{10}{\degree}$, such that the flow separates at approximately 12\% chord with the specified flow parameters~\citep{Xu2023}. The SJAs used are the commercially available Murata MZB1001T02 microblowers which are embedded underneath the surface of the wing model. The array consists of two rows of 12 SJAs located at 10.7\% and 19.8\% chord, however, only the upstream row was used in this experiment. The SJA was operated near its resonant frequency with a square-wave carrier frequency of $f_c=25.5$~\unit{\kilo\hertz}. The SJA was burst modulated at an excitation frequency of $200$~\unit{\hertz}, corresponding to a nondimensional frequency of $F^+=11.76$, where $F^+=f_mc/U_\infty$. The blowing strength of SJAs was varied by operating them at three voltages: 5, 10, and 20 $\mathrm{V}_{pp}$, corresponding to  blowing ratios of $C_B=0.47$, 1.5, and 3.8, respectively~\citep{Xu2023}.

Smoke visualization was achieved with a horizontal wire, coated manually in oil, and heated resistively. The smoke wire was positioned upstream of the leading edge, along the chordline, so that the smoke streaks followed the edge of the shear layer~\citep{Machado2023}. Images were captured with a Nikon D7000 DSLR with long exposure times (1/4~\unit{\second} and 1/6~\unit{\second}) to provide a sense of the mean flow. A continuous laser line generator was used to accurately illuminate the smoke streaks. For the overhead flow visualizations, the laser line was passed through a diverging lens, positioned downstream of the test section, which provided a volume of illumination above the airfoil in Fig.~\ref{fig:smoke visualization method}. The smoke streaks were then imaged with the camera above the wind tunnel, through an acrylic observation window. Sectional visualizations of the shear layer at the trailing edge were achieved with the laser sheet oriented perpendicular to the flow, along the trailing edge, and the camera downstream of the test section.

\begin{figure}
    \centering
    \includegraphics[width=\linewidth]{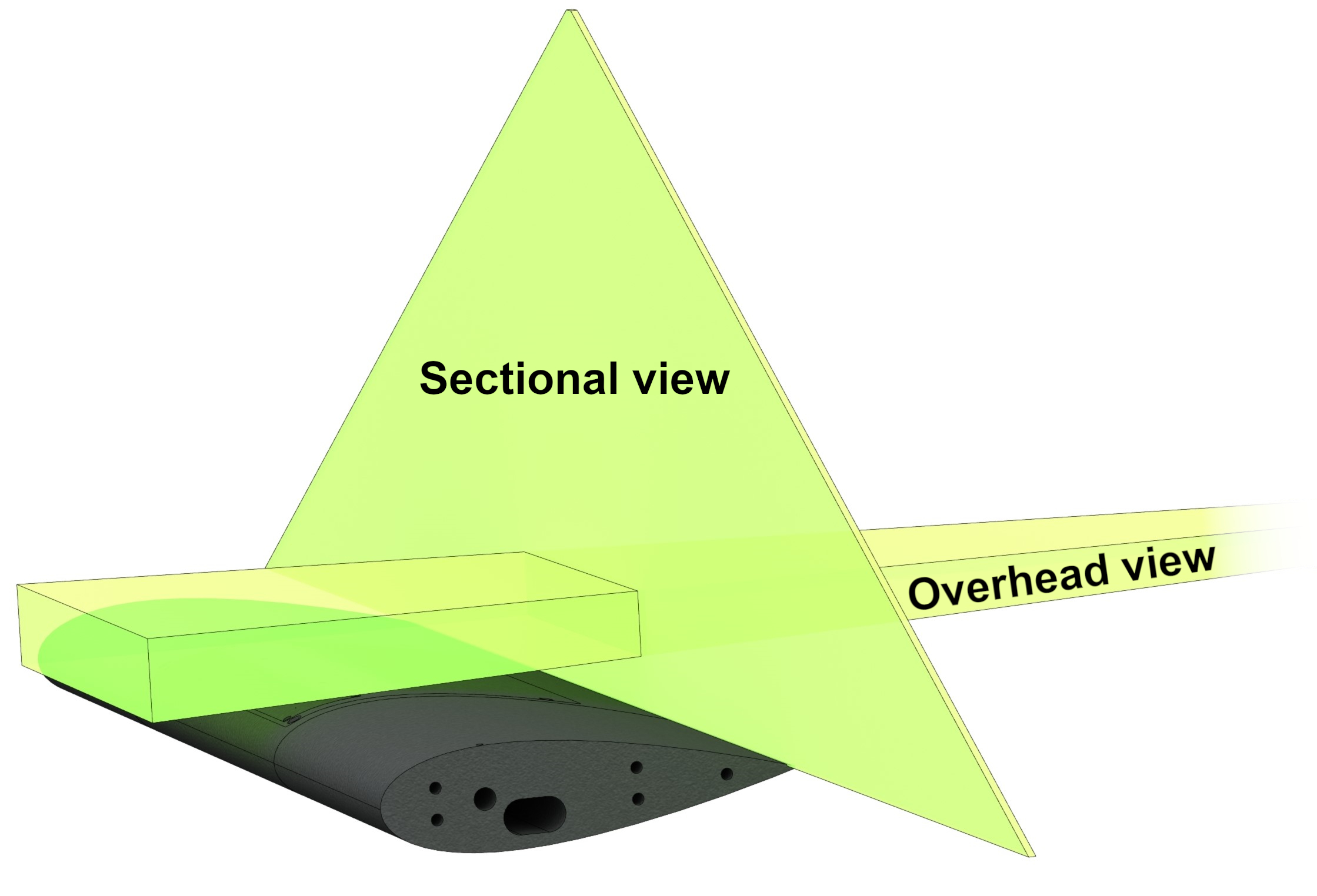}
    \caption{Depiction of the airfoil model and laser orientations for the overhead and sectional flow visualizations}
    \label{fig:smoke_visualization_method}
\end{figure}

\section*{NUMERICAL METHOD} %make d in $$ for italics, and use \unit or \SI
For the numerical simulations, a duct cross-flow geometry was modelled, this geometry was chosen as it allows the SJA effect to be more distinct and identifiable. The primary numerical computations were performed in OpenFOAM~\citep{Jasak2009} with 3D Unsteady Reynolds-Averaged Navier-Stokes (URANS) simulations using the Launder-Sharma Low-Reynolds number $k-\epsilon$ model~\citep{Launder1974}. To match the experimental boundary conditions, a similar characteristics synthetic jet actuator (SJA) with matching jet momentum and modulated frequency was modelled and validated~\citep{Feero2015}. The finite volume method based governing equations are solved with the a PISO algorithm based solver and the convergence criterion of $10^{-6}$ is assigned for all variables, a Courant number of unity is maintained by an adjustable time step $\Delta t$.

\begin{figure}[t]
\centering
\includegraphics[width=3in]{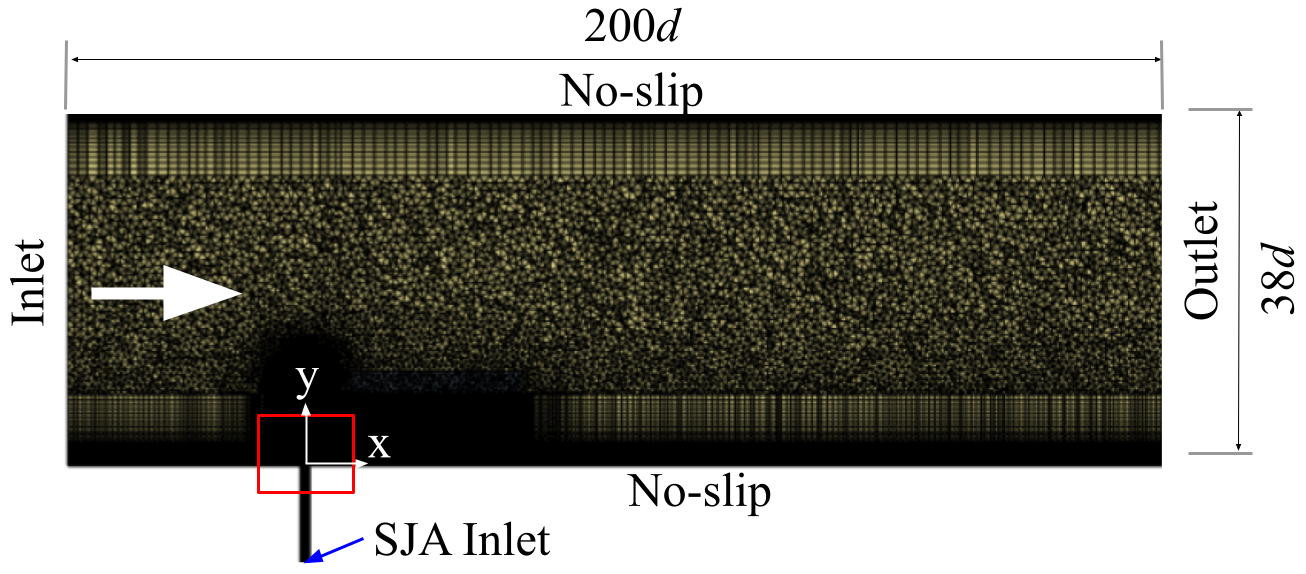}
\caption{Mesh configuration and boundary conditions for the numerical case at symmetry plane (only part of the overall length is displayed).}
\label{fig:mesh}
\end{figure}

The computational domain of the SJA in duct crossflow is presented in Fig.~\ref{fig:mesh}, the SJA jet slot has a diameter $d$ of \SI{2}{\milli\metre} and a neck height of at least $15d$ to satisfy the volume ratio requirement~\citep{ho23}. The crossflow duct domain has a length of $200d$, a height of $38d$ and a width of $20d$, the circular SJA jet slot is located $25d$ downstream of the duct inlet at the symmetry plane. A velocity inlet and pressure outlet boundary condition (BC) was applied to the duct, all walls were assigned no-slip and the symmetry BC was assigned to the front and back of the duct. A uniform velocity inlet of \SI{2}{\metre\per\second} was applied to the duct inlet with an inlet turbulence intensity of 1\%. The analytical Womersley solution was applied to the bottom of the SJA neck to model the effect of the SJA with reduced computational cost \citep{palumbo22,ho23}. Three cases with increasing actuation frequency were simulated under fixed crossflow conditions, Table \ref{Table1}. Sampling for the numerical results began after two flow through cycles ($t^*=t / ( L_{duct}/U_\infty)=2$), the numerical data consisted of 20 samples per cycle over a period of 3 actuation cycles.

\begin{table}[ht]
\centering
\caption{SJA characteristics in numerical cases.}
\label{Table1}
\begin{tabular}{c c c c}
& & \\ % put some space after the caption
\hline
\hline
Case & $f_{c}$ (Hz) & $Re_{\bar{U}}$ & $C_{B}$ \\
\hline
A	& 120	& 1350 & 4.85 \\
B	& 200	& 1340 & 4.82 \\
C	& 425	& 1370 & 4.93 \\
\hline
\hline
\end{tabular}
\end{table}

\section*{EXPERIMENTAL RESULTS}
\begin{figure}
    \centering
    \begin{subfigure}{\linewidth}
        \includegraphics[width=\linewidth]{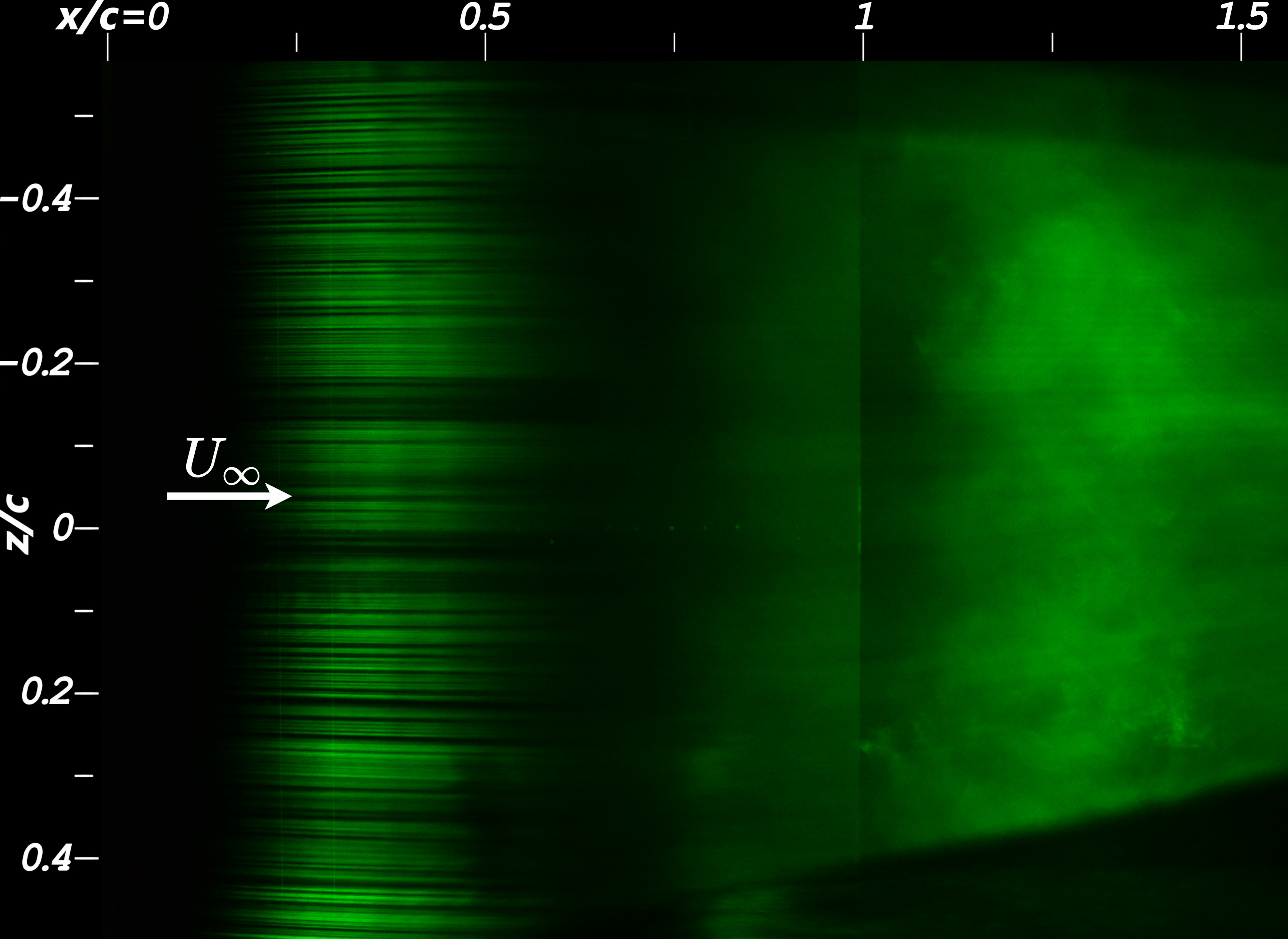}
        \caption{Baseline}
        \label{fig: laser overhead baseline}
    \end{subfigure}
    \begin{subfigure}{\linewidth}
        \includegraphics[width=\linewidth]{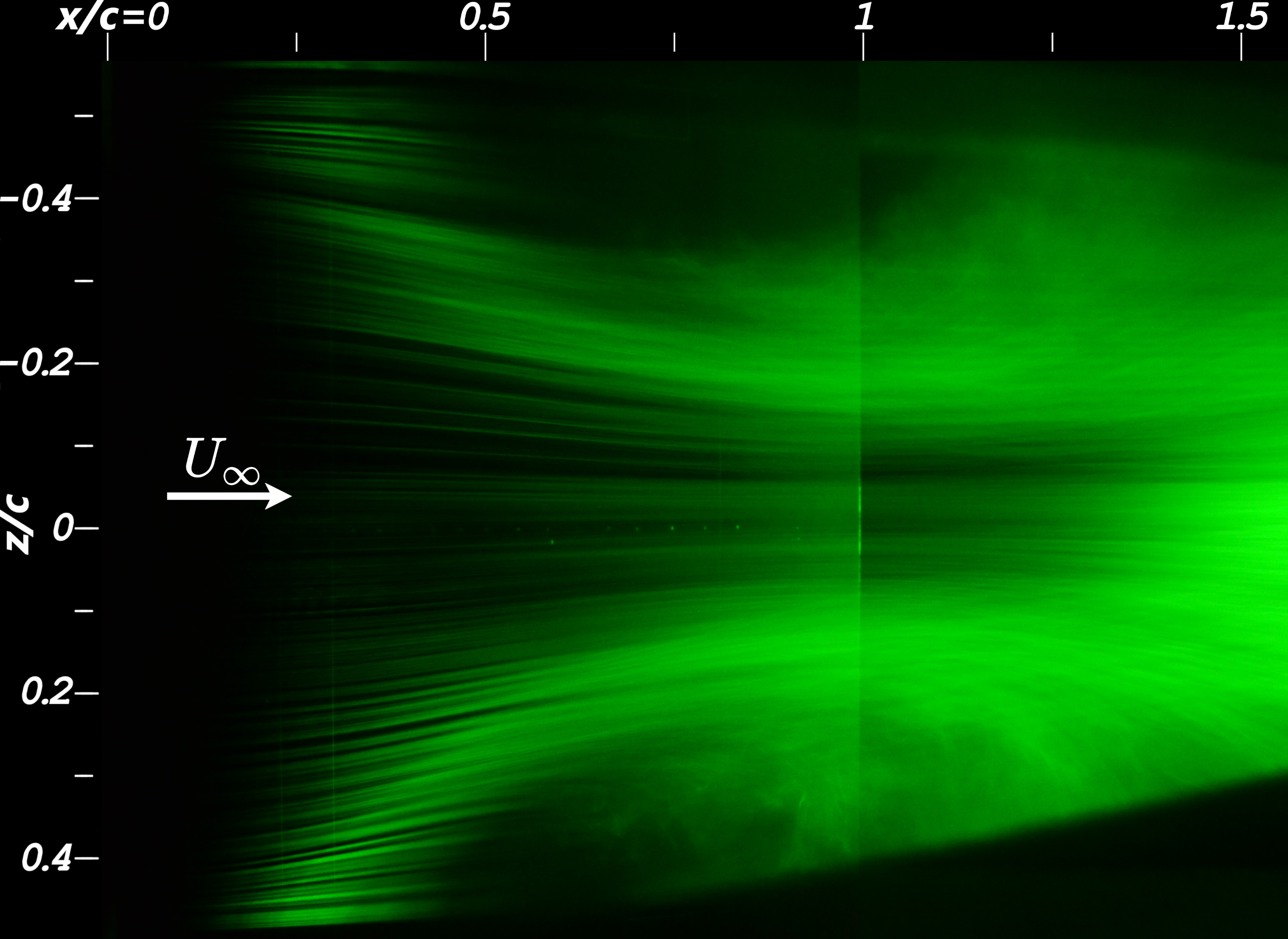}
        \caption{With control at $C_B=3.8$}
        \label{fig: laser overhead 1V}
    \end{subfigure}
    \caption{Overhead smoke visualization of flow above the airfoil}
    \label{fig:laser overhead}
\end{figure}

Overhead visualizations of the mean flow are pictured in Fig.~\ref{fig:laser overhead} for the baseline and control case. In the baseline case, Fig.~\ref{fig: laser overhead baseline}, the streaklines remain laminar up to $x/c=0.6$. Further downstream, the smoke diffuses indicating the transition to turbulence. This result is aligned with a prior study \citep{Machado2023}, which attributed the transition to turbulence to the roll-up of shear layer vortices. The baseline flow is seen to be primarily in the streamwise direction with minimal spanwise velocities. For the control case, Fig.~\ref{fig: laser overhead 1V}, laminar streaklines are observed to persist past the trailing edge, indicating a full reattachment of the flow at the midspan. Additionally, the flow is seen to converge towards the midspan, forming a laminar region between $z/c=\pm0.2$, as evidenced by the smooth streaklines. Beyond this central controlled region, diffuse smoke indicates increased turbulence. The flow contraction was previously studied and attributed to a pressure gradient caused by varying flow speeds across the span, with the fastest fluid at the plane of symmetry of the SJA array~\citep{Machado2023}. Downstream of the trailing edge, the flow diverges again as the wake forms a free shear layer and the pressure gradient diminishes. The effective control region is approximated as only 45\% of the length of the array, highlighting that achieving effective control across the entire span of a wing remains challenging. It is hypothesized that optimizing the SJA parameters could improve the spanwise control authority, namely, the blowing strength, actuation frequency, or the duty cycle. Furthermore, modifications to the model, such as adjusting the number and spacing of the SJAs, their blowing angle, and location, or incorporating multiple rows, might also improve performance.

\begin{figure*}
    \begin{subfigure}{0.325\linewidth}
        \includegraphics[width=\linewidth]{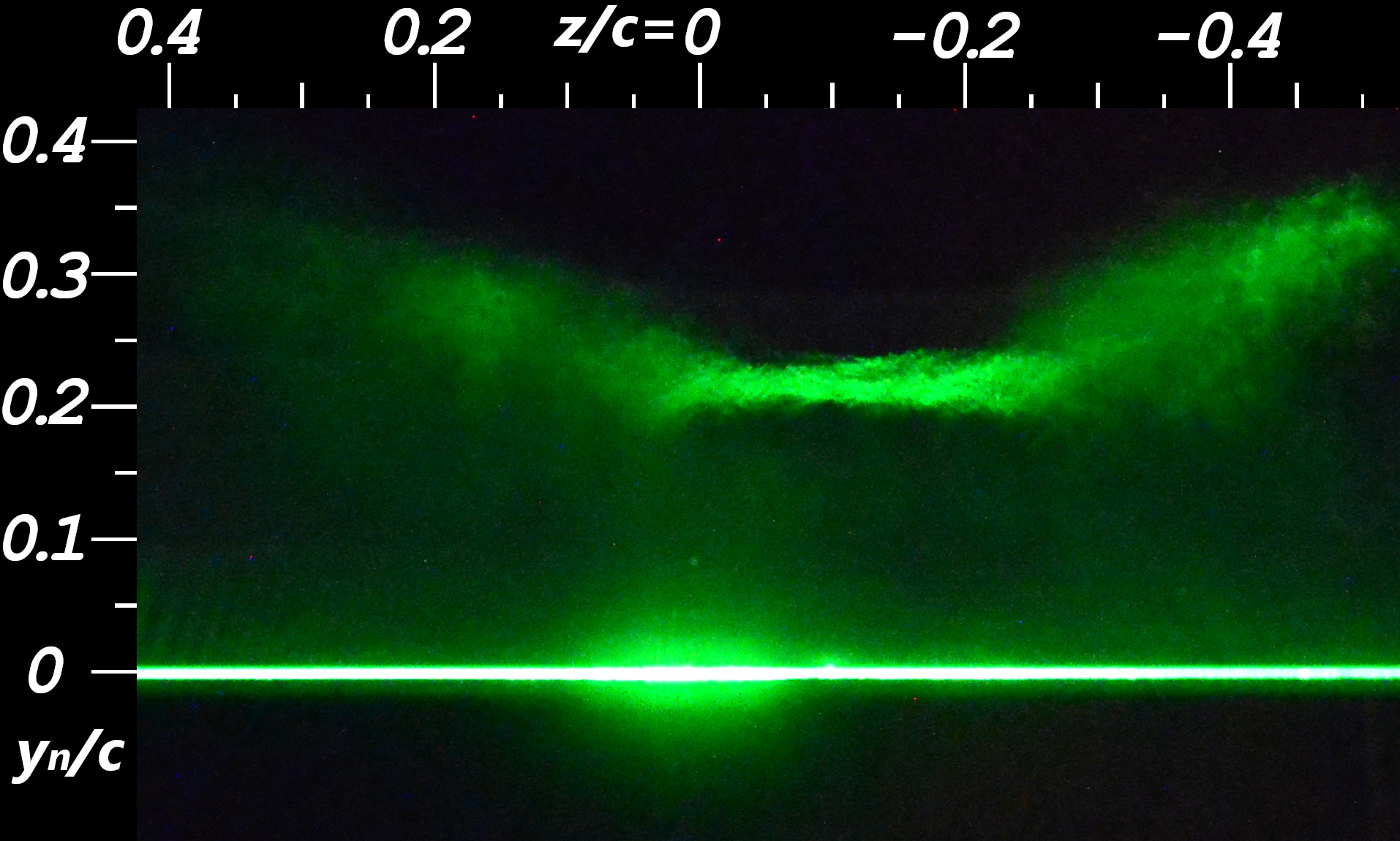}
        \caption{$C_B=0.47$}
        \label{fig:laser_smoke_a}
    \end{subfigure}
    \begin{subfigure}{0.325\linewidth}
        \includegraphics[width=\linewidth]{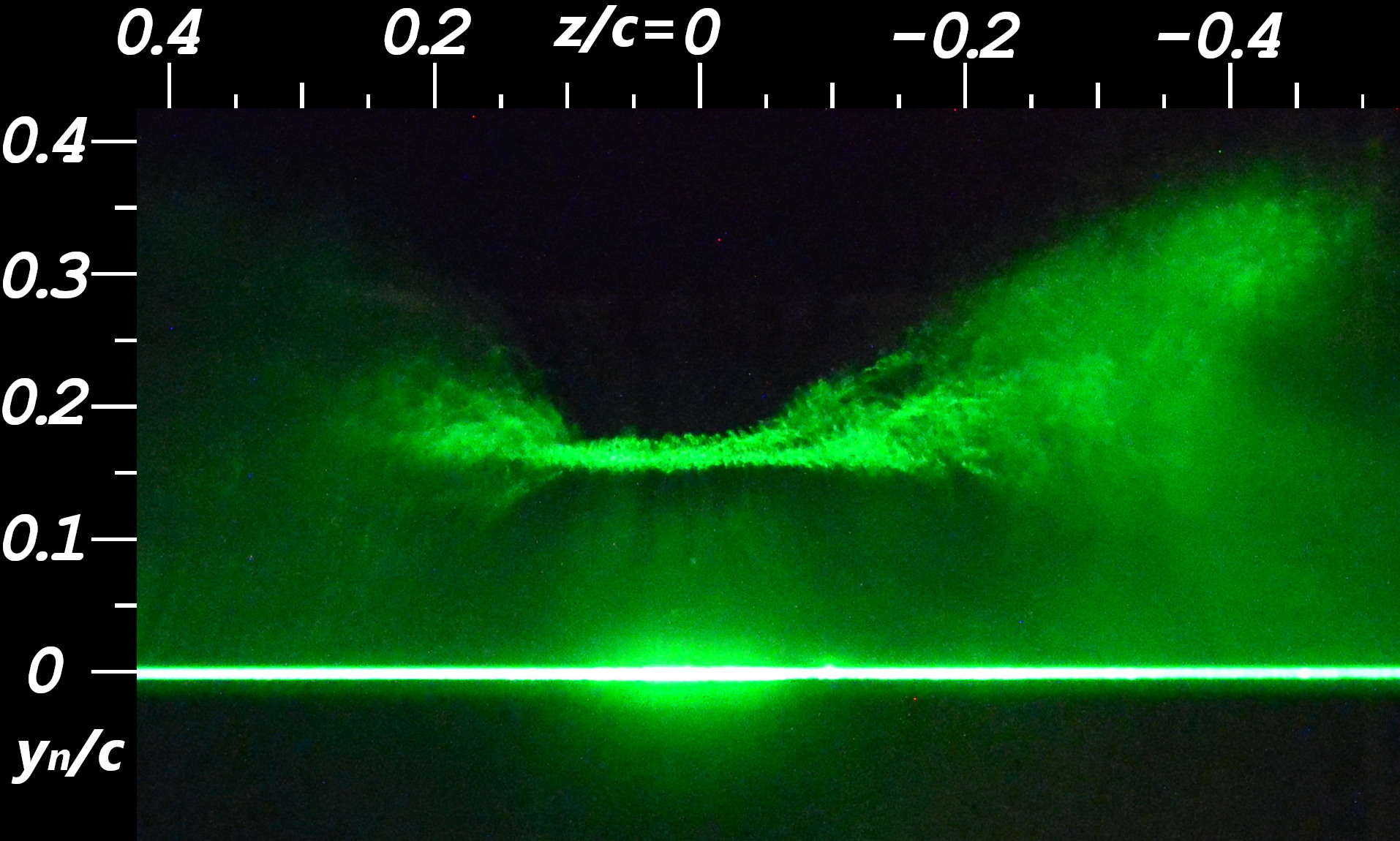}
        \caption{$C_B=1.5$}
        \label{fig:laser_smoke_b}
    \end{subfigure}
    \begin{subfigure}{0.325\linewidth}
        \includegraphics[width=\linewidth]{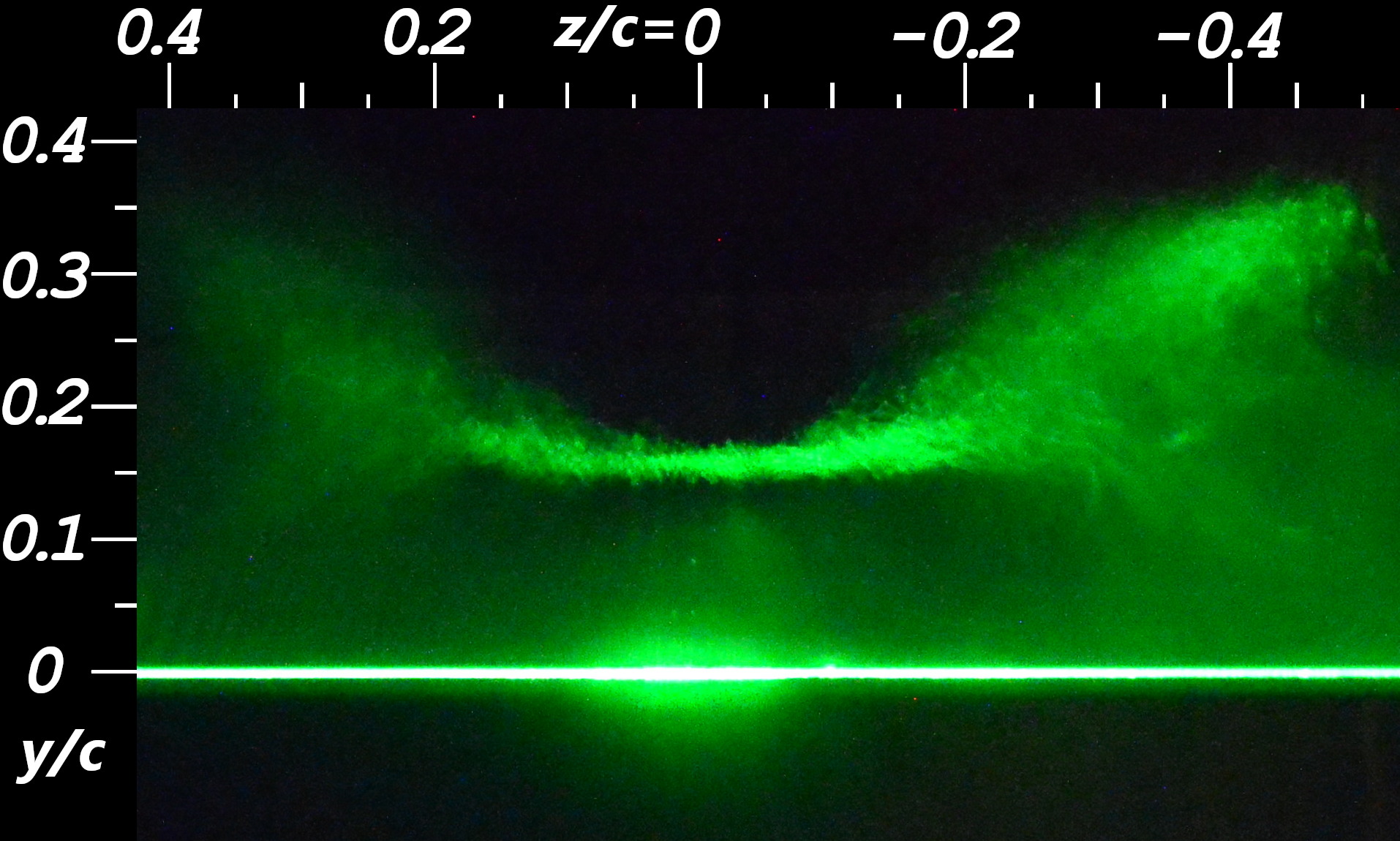}
        \caption{$C_B=3.8$}
        \label{fig:laser_smoke_c}
    \end{subfigure}
    \caption{Sectional smoke flow visualization at the trailing edge; spanwise-transverse plane}
    \label{fig:laser_smoke}
\end{figure*}

Cross-sectional smoke visualizations at the trailing edge are presented in Fig.~\ref{fig:laser_smoke}, providing insights into the shear layer dynamics at three different blowing strengths. At the lowest blowing strength, $C_B=0.47$, the laminar control region is observed to be centered about $z/c=0.12$ instead of the midspan, Fig.~\ref{fig:laser_smoke_a}. This deviation is attributed to the SJAs operating near their lower voltage limit, leading to more variation in jet velocity. Consequently, the SJAs on the right side of the airfoil likely produce stronger jets at the applied voltage, shifting the center of control to the right. Conversely, the flow on the left side of the airfoil is uncontrolled and turbulent, as indicated by the diffuse smoke, suggesting that the jets are not powerful enough to reattach the flow at this spanwise location. An increase in the blowing strength shifts the center of the control region to the midspan, as seen in Fig.~\ref{fig:laser_smoke_b}. Additionally, the shear layer is thinner, as evidenced by the dense streaklines appearing at a lower $y$ location, indicating more effective control. The thinner shear layer, as a result of control at $C_B=1.5$, indicates more effective separation control, resulting in favorable lift characteristics and reduced pressure drag due to a narrower wake. The effective control region begins to break down at $z/c=\pm2$, where the smoke streaks exhibit signs of unsteadiness. A transitional region is observed between the controlled and separated flow regions, characterized by a highly unsteady shear layer that alternates between separated and attached states. An increase in blowing strength to $C_B=3.8$ suppresses the unsteadiness at the edge of the control region, as evidenced by a tighter spread in the streaklines in Fig.~\ref{fig:laser_smoke_c}. Additionally, the transition area is less prominent; instead, an abrupt change in flow characteristics occurs at the edge of the control region. The shear layer at the midspan appears identical in thickness and steadiness to that observed in Fig.~\ref{fig:laser_smoke_b}, indicating that the midspan control effects are saturated at $C_B=1.5$. However, the spanwise extent, particularly the steadiness, of the controlled flow continues to improve.

\section*{NUMERICAL RESULTS}
\begin{figure}
\centering
\includegraphics[width=3in]{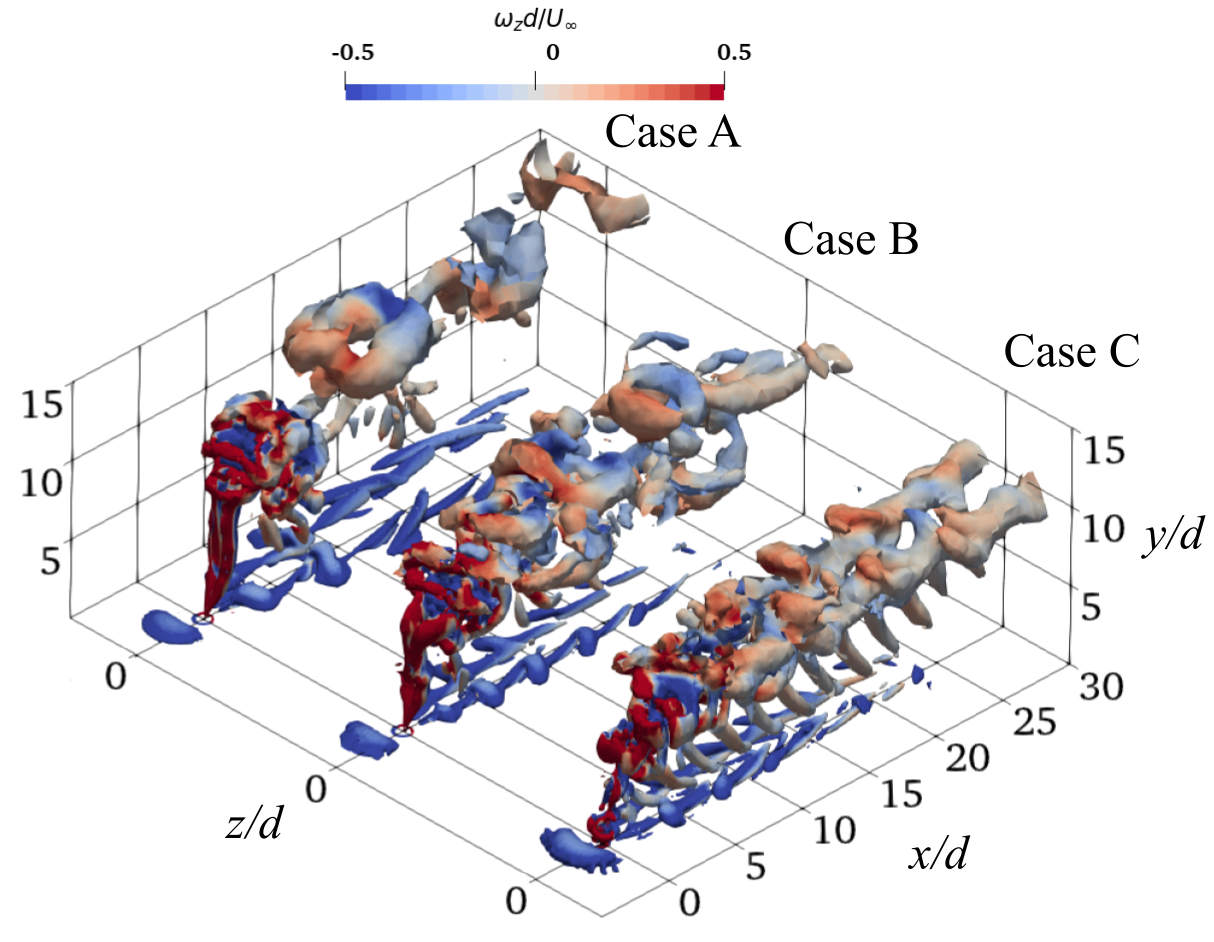}
\caption{Isometric view of instantaneous Q-Criterion contours at t = 0.5s, coloured with normalized spanwise vorticity (Threshold: $Q^*=Qd^2/U_\infty ^2 = 0.01$).}
\label{fig:q-cri-iso}
\end{figure}

To identify the 3D vortical structures of a discrete SJA, the isosurface of Q-criterion contours are generated at $t^*=2.5$. The isometric view of the Q-criterion contours are presented in Fig.~\ref{fig:q-cri-iso}, coloured by spanwise vorticity. The effect of the actuation frequency is immediately noticed based on the variation of the different structures and density formed in the crossflow. At the lowest actuation frequency of Case A, a distinct tilted vortex ring is formed, convecting downstream rapidly \citep{ho22,jabbal2010}. In Case B and C, the expelled vortex ring is less visible due to the more complex flow structures, a result of the increased flow mixing from higher actuation frequency. At the highest frequency of Case C, the formation of a tail like structure was observed along the symmetry plane. This tail like structure is tiled towards the downstream direction and significantly less prominent at lower frequency cases. This tail like structure also differs from the legs of a hairpin vortex or the trailing vortex pair of a tilted vortex ring, as it is tiled towards the downstream direction, rather than towards the SJA jet slot exit. This behaviour is likely due to the significant blowing ratio $C_B$ and the low thickness of the boundary layer, where the structure is subjected to a more uniform crossflow than cases with longer boundary layer development length. The increase of actuation frequency also influenced the near wall tertiary vortices, the streamwise distance between each vortex has an inverse relation with the SJA frequency, as expected. However, the lowest frequency of Case A resulted in wider near wall tertiary vortices in the spanwise direction than the other two cases.

\begin{figure}
\centering
\includegraphics[width=3in]{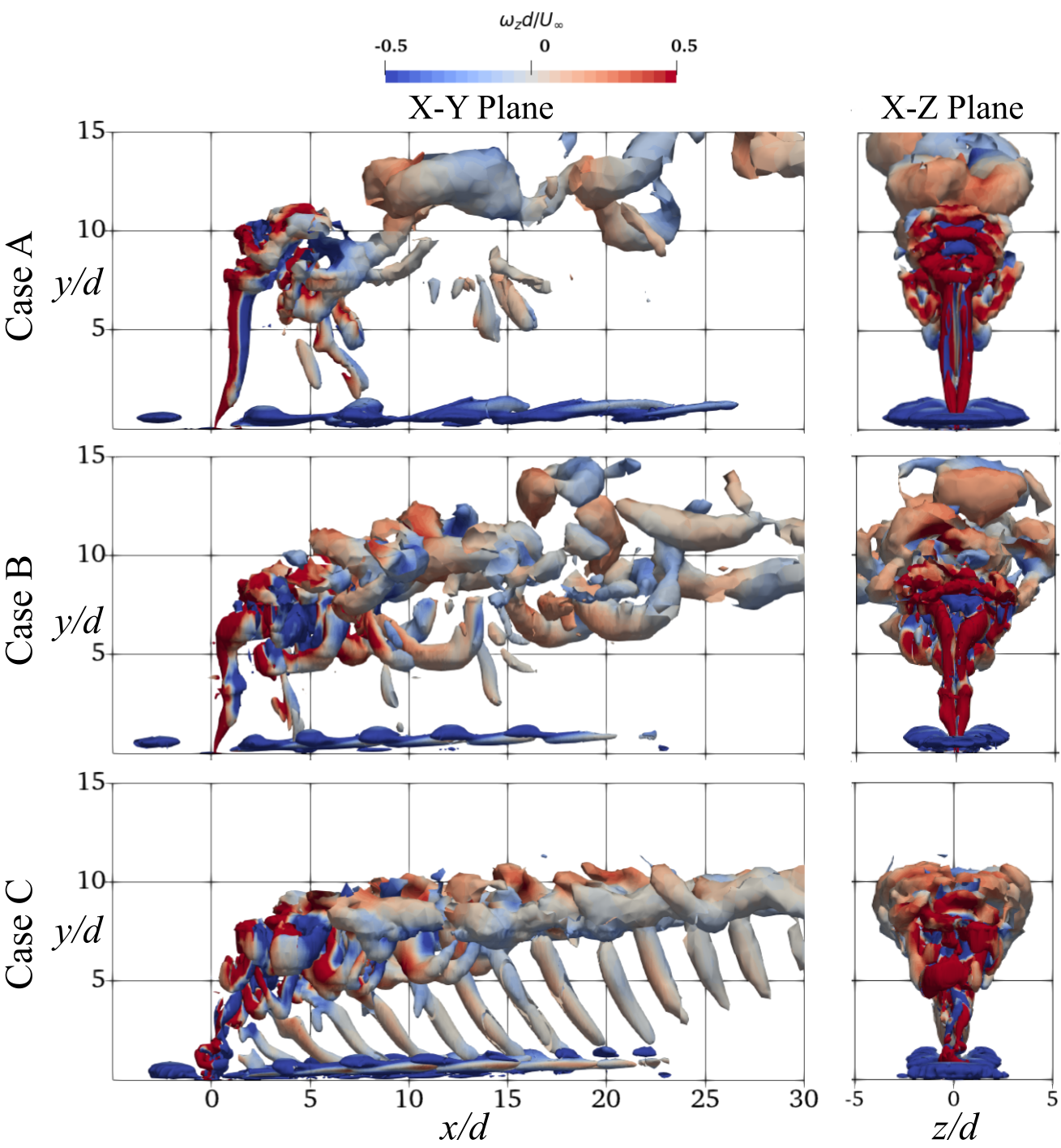}
\caption{Side and front view of instantaneous Q-Criterion contours at $t*=2$, coloured with normalized spanwise vorticity (Threshold: $Q^*=Qd^2/U_\infty ^2 = 0.01$).}
\label{fig:q-cri-side}
\end{figure}

The effect of the actuation frequency was further examined with side and front view of the Q-criterion contours in Fig.~\ref{fig:q-cri-side}. Case A with the lowest frequency had the greatest boundary layer penetration, this is due to the lower frequency allowing greater momentum per cycle into the crossflow. As the actuation frequency increases, the SJAs still shared the same time-averaged jet momentum, but it is spread out into more cycles and thus less powerful per cycle. The overall spanwise effect of SJAs appeared to be modest, the primary expelled structure were within $5d$ form the symmetry plane in all cases. The near wall tertiary vortices were even more limited in the spanwise direction, it was observed to only extend up to $3.5d$ from the symmetry plane and narrows as the actuation frequency was increased.

\begin{figure}
\centering
\includegraphics[width=3in]{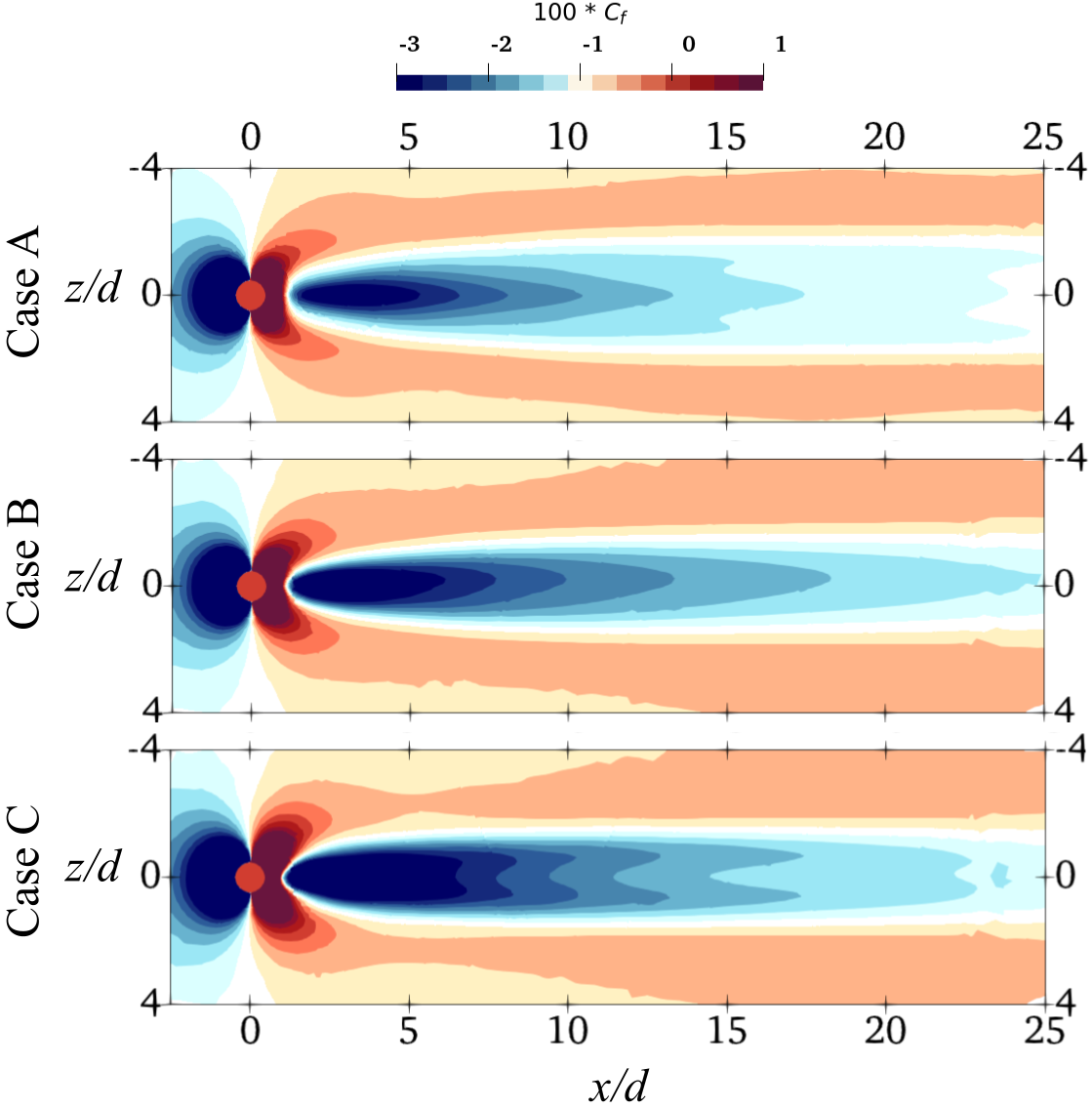}
\caption{Time-averaged skin friction coefficient contour comparison
of the lower wall.}
\label{fig:Cf}
\end{figure}

To further examine the impact of the actuation frequency and near wall tertiary vortices on the potential in flow separation control, time-averaged skin friction coefficient ($C_f$) contours were generated in Fig.~\ref{fig:Cf}. Upstream of the jet slot exit, a region of increased $C_f$ was observed, this is contributed by the ingestion cycle of the actuation, drawing fluid into the SJA, resulting in greater near wall flow velocity. While the upstream $C_f$ appeared near identical across all three cases, the downstream side of the SJA is affected noticeably by the actuation frequency. An overall increase of $C_f$ along the lower wall was observed downstream of the jet exit with the increase of actuation frequency. Case C had the narrowest near wall tertiary structures among all three cases as observed in Fig.~\ref{fig:q-cri-side}, but the $C_f$ spanwise coverage displayed the opposite effect. The $C_f$ streamwise distribution of Case C also differed from the other two cases, the profile between $5<x/d<20$ resembles that of the hairpin vortex with the twin peak. This change of behaviour is likely a result from the tail like structure along the symmetry plane, decreasing the symmetry plane near wall flow velocity and $C_f$.

\section*{CONCLUSIONS}
In this study, an array of SJAs was employed to reattach the flow over a stalled NACA 0025 airfoil experimentally, with a focus on understanding the three-dimensional characteristics of the controlled flow. The visualizations revealed that the controlled flow contracts towards the midspan, forming a central control region with a spanwise length of approximately 0.4$c$, indicating that the spanwise extent of control is limited to only 45\% of the SJA array's span. Sectional smoke visualizations provided insights into the shear layer behavior across the span at the trailing edge, highlighting the effect of the momentum coefficient. The results indicate that the spanwise control authority continues to improve even after the midspan control effects are saturated. The presence of a transitional region between the fully attached and separated zones was identified, characterized by an unsteady, flapping shear layer. Additionally, control near the lower end of the microblower's operating voltage revealed that variations in the synthetic jet velocity can lead to asymmetrical control effects. A complementary numerical simulation was performed on three different SJAs in a duct crossflow, the same blowing ratio matches that of the experiment and the actuation frequency was varied. It was observed that the increase of actuation frequency resulted in more compact structures with reduced boundary layer penetration. A change in crossflow structures was also observed at the highest frequency case, displaying a tail like structure that tilts towards the downstream direction. The spanwise control potential of all three cases appeared modest especially at near wall region, with a maximum spanwise coverage of $5d$ from the symmetry plane. An overall increase of skin friction was observed downstream of the jet slot exit with the actuation frequency, with a change of profile at the highest frequency case.

\bibliographystyle{tsfp}
\bibliography{tsfp}

 \end{document}